\documentstyle[twocolumn,prc,aps,epsf,epsfig]{revtex}
 \begin{document}
 
\newcommand{\be}{\begin{eqnarray}}
\newcommand{\ee}{\end{eqnarray}}
\twocolumn[\hsize\textwidth\columnwidth\hsize\csname@twocolumnfalse\endcsname
\title{Possible evidence 
for ``dark radiation'' from 
 Big Bang Nucleosynthesis Data
}

\author{ V.V. Flambaum$^1$ and E.V. Shuryak$^2$ }


\address{$^1$
 School of Physics, The University of New South Wales, Sydney NSW 2052,
Australia and Physics Division, Argonne National Laboratory, Argonne,
 Illinois 60439-4843, USA}

\address{$^2$ 
Department of Physics and Astronomy, State University of New York, 
Stony Brook NY 11794-3800, USA
}

\date{\today}
\maketitle

\begin{abstract}
We address the emerging discrepancy between the Big Bang 
Nucleosynthesis data and standard cosmology, which asks for 
a bit longer evolution time. If this effect is real,
  one possible implication (in a framework of brane cosmology model)
is that there is a ``dark radiation'' component which is negative and
makes few percents of ordinary matter density. If so, all scales
of this model can be fixed, provided brane-to-bulk leakage problem
is solved.
\end{abstract}
\vspace{0.1in}
]
\begin{narrowtext}
\newpage



  In this paper we discuss
the emerging  discrepancy between
Big-Bang Nucleosynthesis (BBN) calculations and measurements.
The facts themselves are rather well known. For example, in
 recent summary 
of the BBN data analysis can be found e.g. in 
\cite{Barger:2003zg} who conclude that ``effective number of extra 
neutrinos'', as it is usual to put it, does not want to go away and
tends to be negative. The final numbers these authors give
is 
\be \label{eqn_Steigman} \Delta N_\nu=-0.65\pm 0.35 \ee
Account for other forms of matter naturally make the discrepancy
larger,
for example see
 the effects 
 of right-handed neutrinos \cite{Dolgov}.

 Thus, formally speaking, the effect is
about two standard deviations, and in principle one may not
worry about it yet.
However, in this letter we propose to take another attitude,
consider it to be real effect and discuss what it potentially may mean,
either for physics or cosmology, which we now discuss subsequently. 

 {\bf Variable physical constants} is one possibility to explain the
 anomaly. One may take the following attitude: if the deviations
are not some systematics in part of the data but a real effect,
it should be possible to find its origin by letting main
parameters of the BBN program to be fitted freely, unconstrained to
the today's value, and see if it results in much better fit of
all the data. This was found indeed to be the case, and
the best results were obtained if the varied quantity is
the {\em deuteron binding} $Q_d$ \cite{DF_un}. Its variation by
\be
\delta Q_d/Q_d=-0.019\pm 0.005
  \ee
leads to much better fit of all the same data (and thus
the effect makes larger
number of standard deviations). On top of that, this change
leads to excellent agreement
for baryon-to-photon ratio $\eta$ between this fit and the best
fit to Cosmic Microwave Background (CMB) data.

The deuteron binding is not just one of the parameters; in fact it
was identified in our papers \cite{FS} as part of the chain
which is the most sensitive way to test
possible variation of weak-to-strong scales. The
above mentioned variation of $Q_d$ corresponds to only
 $10^{-3}$ variation of the strange quark to QCD scale\footnote{Or, if
  one prefers, a
  twice smaller variation of the kaon to nucleon mass ratio.}
  $ \Delta (m_s/\Lambda_{QCD})/(m_s/\Lambda_{QCD})$.
The description of the corresponding enhancement factors which lead
all the way from this small variation to that of deuteron binding and
eventually to quite noticeable variation of $He^4,Li^7$
production can be found in these papers.

{\bf Deviations from standard cosmology\footnote{By standard we mean
    textbook cosmology with matter consisting of photons and
 3 massless neutrinos.}}. 
One may try to ascribe the discrepancy to
an {\em increase}
of the total time
allocated to BBN nuclear reactions
 (e.g. more neutrons should decay).

The first obvious suspect is the cosmological scalar field
(quintessence \cite{quint}), which may or may not be related to
either inflation-era scalar or to present-day acceleration,
see discussion e.g. in \cite{Barger:2003zg}.
But, if so, one would need to have an opposite sign of its
effective potential, which is quite difficult to reconcile
with other constraints.

 {\bf Brane Cosmology} much discussed
in literature is another attractive option.
 We restrict this discussion to the simplest
version of it, with a single brane in multi-dimensional
space.

The unavoidable consequence of gravity propagating in extra
dimension is a sort of  back reaction on the 
cosmological evolution as observed on the brane. 
New terms due to second order effects in density and
due to bulk gravity field  appear 
in effective evolution equation on the brane,
see \cite{Shiromizu:1999wj}
and a review in \cite{Kiritsis:2003mc}

The parameters of the model include the 4-d brane tension
$\lambda=M_\lambda^4$,
 the 5d cosmological constant $\Lambda$,
  and the
5d gravity coupling $\kappa_5^{-2}=M^3_G$. One relation between
them is need to be fine-tuned to get effective cosmological
constant to a small value of $\lambda_4$ 
(to be ignored at BBN time we discuss).
Another relation is to fix the Newton constant
 $G_N=\kappa_5^4\lambda/48\pi$. That leaves us with one free
 parameter
  to be fixed.  

The resulting evolution equations of the model can be written
(see review
\cite{Kiritsis:2003mc} and references therein)
 in the following convenient form
\be
\dot{\rho}+3(1+w)\,{{\dot a}\over a} \, \rho = -T^{leak} \\
 \dot\chi+4\,{{\dot a}\over
a}\,\chi= T^{leak}
\label{rho1}
\ee
\be
{{{\dot a}^2}\over {a^2}}=A\rho^2+{8\pi\over 3 M_P^2}(\rho+\chi) -
{k\over{a^2}}+\lambda_4
\label{a1}
\ee
where $M_P$ is the Plank mass and
 $w=p/\rho$ is the usual matter equation of  state.
The dots stand for derivatives with respect to
 cosmological time $\tau$. $k$-term (due to curvature) and
 $\lambda_4$ are unimportant. The coefficient $A$ of the quadratic 
term in density is  suppressed
compared to the linear term
by very small parameter $T^4/M_\lambda^4$ where $T^4\sim \rho$ is the
temperature scale. The ``dark radiation'' term $\chi$ 
is the leading correction to standard cosmology, and it has the same
evolution as radiation-dominated matter $w=1/3$ (appropriate at both BBN and
CMB times), and the term $T^{leak}$ describes ``leakage'' of energy
from the brane to the bulk which happens at early (noninear)
stage. Note
that $\rho+\chi$ is conserved.

Let us translate the BBN discrepancy in terms of $\chi$
\be \label{eqn_chitorho}
{\chi\over \rho}={7\over 43}{\Delta N_\nu\over N_\nu} \approx -0.11\pm 0.06 \ee
where the number is for (\ref{eqn_Steigman}).

For expanding brane universe, it is quite natural
\cite{Kiritsis:2003mc} to obtain
negative $\chi$. The specific mechanisms of the ``leakage''
depend on details of the nonlinear evolution stage, for some
model-dependent  recent calculations see \cite{LS}.
Note that our results are consistent with the limits on  $\rho/\chi$
which have been obtained earlier \cite{chi}.

{\bf Implications.}

The main consequence one can draw from possible observation
of ``dark radiation'', if confirmed,
 would be a definite magnitude for 
the ``leakage'' term, potentially capable to fix the
remaining scale ambiguity of the model. 

Note that the number of effective degrees of freedom in a Standard
Model is reduced by about an order of magnitude, from early to BBN
time. Therefore, $\chi$, when produced, was even smaller correction 
to the ordinary matter density.
 
Another implication\footnote{Unlike models with
  scalar fields, 
 the  ``dark radiation''  has rigidly fixed
   time evolution.}
is because that both BBN and CMB times
are in the radiation-dominated era, and therefore the same ratio
$\chi/\rho$ should be kept.
As detailed in \cite{Barger:2003zg},  the current data
on CMB do not contradict the same value of $\Delta N_\nu$,
 although their
accuracy is less restrictive at this time.
 With better data on both BBN and CMB  one can possibly test
whether both will give the same value of  $\chi/\rho$. If not,
the brane cosmology model may thus be rejected.

Suppose we take this literally and ask what implications
this would have for earlier cosmology. Unlike the usual matter,
which presumably was produced at the end of inflation era at some
particular time, the ``dark radiation'' was produced
gravitationally, perhaps much earlier, and its amount
depends on such parameters as the number of extra dimensions.
With a non-zero $\chi$, one should seriously revise  inflation scenarios.

      
 VF appreciates support from the Australian Research Council,
 Department of Energy, Office of Nuclear Physics, Contract No.  W-31-109-ENG-38
 and State University of New York. We are grateful to Ken Nollett for useful
comments.

\end{narrowtext}

\begin{thebibliography}{99}


\bibitem{Barger:2003zg}J.~P.~Kneller and G.~Steigman,
  Phys.\ Rev.\ D {\bf 67}, 063501 (2003)
  [arXiv:astro-ph/0210500].
  V.~Barger, J.~P.~Kneller, H.~S.~Lee, D.~Marfatia and G.~Steigman,
  Phys.\ Lett.\ B {\bf 566}, 8 (2003)
  [arXiv:hep-ph/0305075].


\bibitem{Dolgov}
 A.D. Dolgov,Phys.Rept.370:333-535,2002; hep-ph/0202122

\bibitem{DF_un} V.~F.~Dmitriev, V.~V.~Flambaum and J.~K.~Webb,
  Phys.\ Rev.\ D {\bf 69}, 063506 (2004)
  [arXiv:astro-ph/0310892].



\bibitem{FS}
V.~V.~Flambaum and E.~V.~Shuryak,
Phys.\ Rev.\ D {\bf 67}, 083507 (2003)
[arXiv:hep-ph/0212403].
Phys.\ Rev.\ D {\bf 65}, 103503 (2002)
[arXiv:hep-ph/0201303].


\bibitem{quint} B.Ratra and P.J.E.Peebles, Phys.Rev.D37,3406 (1988),
C.Wetterich,Nucl.Phys.B302, 668 (1988).


\bibitem{Shiromizu:1999wj}
  T.~Shiromizu, K.~i.~Maeda and M.~Sasaki,
  Phys.\ Rev.\ D {\bf 62}, 024012 (2000)
  [arXiv:gr-qc/9910076].

\bibitem{Kiritsis:2003mc}
  E.~Kiritsis,
  Fortsch.\ Phys.\  {\bf 52}, 200 (2004)
  [Phys.\ Rept.\  {\bf 421}, 105 (2005)]
  [arXiv:hep-th/0310001].

\bibitem{LS} D.Langlois and L.Sorbo, Phys.Rev.D68 (2003) 084006, hep-th/0306281
  M.Minamitsuji, M.Sasaki  and  D.Langlois, gr-qc/0501086

\bibitem{chi} K. Ichiki, M. Yahiro, T. Kajino, M. Orito, and G.J. Mathews.
Phys. Rev D {\bf 66}, 043521 (2002); astro-ph/0203272. J.D. Bratt, A.C. Gault,
 R.J. Scherrer, T.P. Walker. Phys. Lett. B{\bf 546}, 19 (2002).
    
\end{thebibliography}
\end{document}